\documentclass[aps,prl,twocolumn,groupedaddress,showpacs]{revtex4}
\pdfoutput=1
\usepackage{graphicx}

\begin{document}

\title{How do Quasicrystals Grow?}

\author{Aaron S. Keys$^1$}
\author{Sharon C. Glotzer$^{1,2}$}
\affiliation{$^1$Department of Chemical Engineering and $^2$Department of Materials Science and Engineering \\University of Michigan, Ann Arbor, Michigan 48109-2136}

\date{\today}

\begin{abstract}
Using molecular simulations, we show that the aperiodic growth of quasicrystals is controlled by the ability of the growing quasicrystal `nucleus' to incorporate kinetically trapped atoms into the solid phase with minimal rearrangement.  In the system under investigation, which forms a dodecagonal quasicrystal, we show that this process occurs through the assimilation of stable icosahedral clusters by the growing quasicrystal.  Our results demonstrate how local atomic interactions give rise to the long-range aperiodicity of quasicrystals.
\end{abstract}

\maketitle

Quasicrystals\cite{shecht84} are a unique class of ordered solids that display long-range aperiodicity, which distinguishes them from ordinary crystals.  It is not known what `special' qualities systems must possess in order to form quasicrystals versus crystals.   Quasicrystals, like crystals, form via nucleation and growth\cite{shecht84}, where a microscopic `nucleus' of the solid phase spontaneously arises in the supercooled liquid and spreads outward, converting the system from liquid to solid\cite{kelton91}.  A fundamental puzzle in quasicrystal physics is to understand how the growth phase of nucleation and growth can lead to a structure with long-range aperiodicity.  Quasicrystals cannot grow like crystals, where the nucleus surface acts as a template for copying a unit cell via local interactions.  Rather, quasicrystals, require specialized ``growth rules" that dictate their formation\cite{grimm02}.

Quasicrystal (QC) growth rules fall into two categories: energy-driven quasiperiodic tiling models\cite{levine84, jeong97} and entropy-driven random tiling models\cite{elser85, ox93}.  While energy-driven models rely on ``matching rules'' to dictate how atomic clusters or tiles attach to the nucleus, entropic models allow tiles to attach randomly to the nucleus with some probability.  Although these models provide important insight into how QCs might form, the physical driving force underlying QC growth, and whether it is based on local interactions or long-range correlations, is not well understood.

In this section, we elucidate the physical mechanism underlying QC growth by studying the post-critical irreversible growth of a metastable dodecagonal QC from a simulated supercooled liquid.  We show that QC growth is facilitated by structurally persistent atoms in low energy motifs that become kinetically trapped in their local configurations in the region surrounding the solid nucleus.  As the nucleus grows, it incorporates these atoms in a way that minimizes expensive rearrangements and hastens solidification, allowing the QC to form instead of the stable crystalline approximant phase.  In the system under investigation, we find that structurally persistent atoms were in icosahedral clusters prior to attaching to the nucleus.  Our results demonstrate how the long-range aperiodicity of QCs arises from local atomic interactions, thus providing a significant step forward in understanding the origin of the QC state. 

\begin{figure}
\includegraphics[width=3.75in]{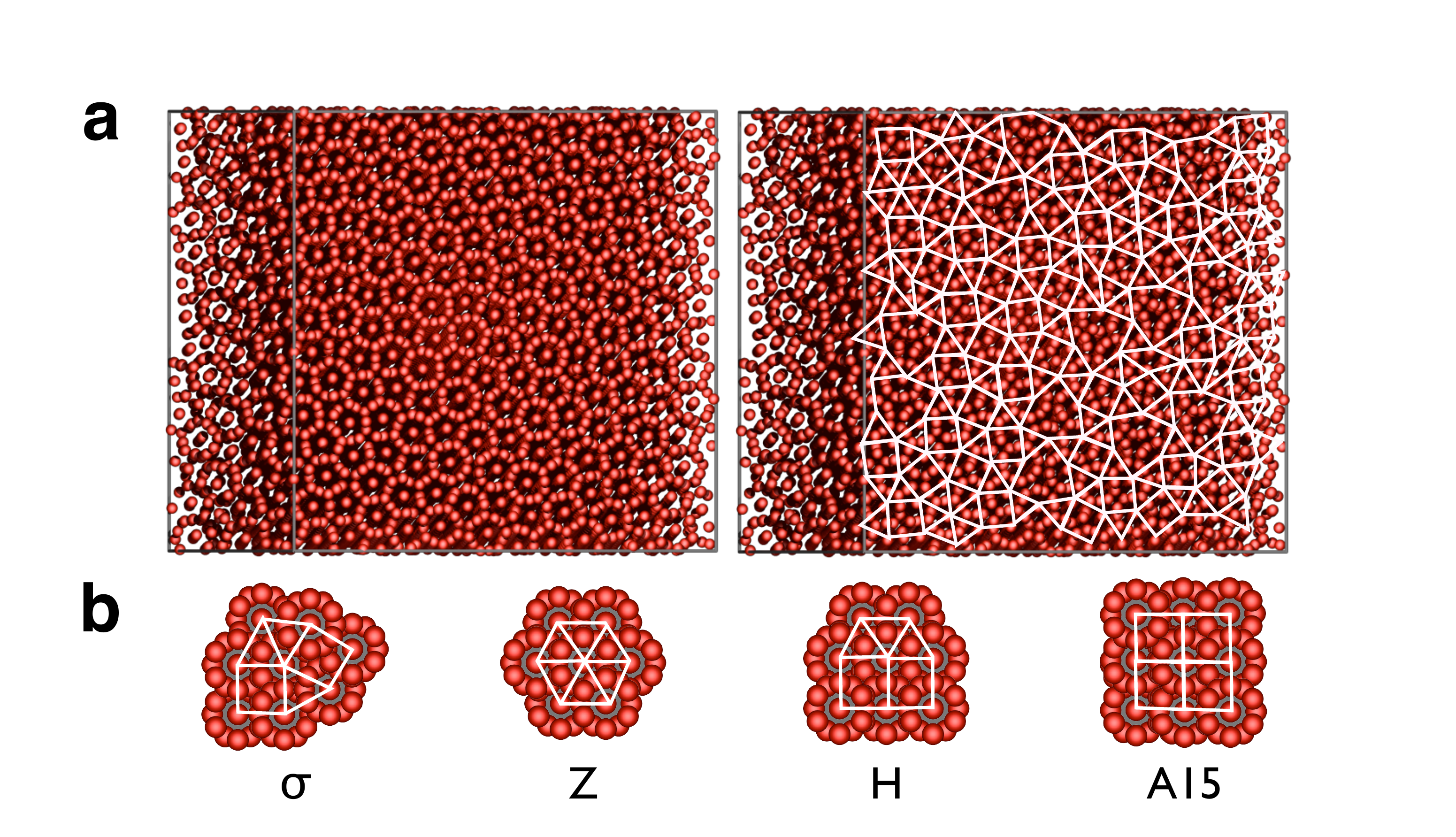}
\caption[Dzugutov Dodecagonal Quasicrystal and Approximants]{Dodecagonal QC and approximants.  (a) 17,576 atom dodecagonal QC formed by the Dzugutov system using molecular dynamics at T=0.42 and $\rho$=0.85, instantaneously quenched to T=0.  The image on the right shows the aperiodic tiles formed by connecting the centers of the dodecagonal rings of atoms.  (b)  Unit cells of various QC approximants.}
\label{fig:aFigure1}
\end{figure}

\begin{figure*}
  \begin{center}
    \begin{minipage}[m]{0.71\linewidth}
      \includegraphics[width=\linewidth]{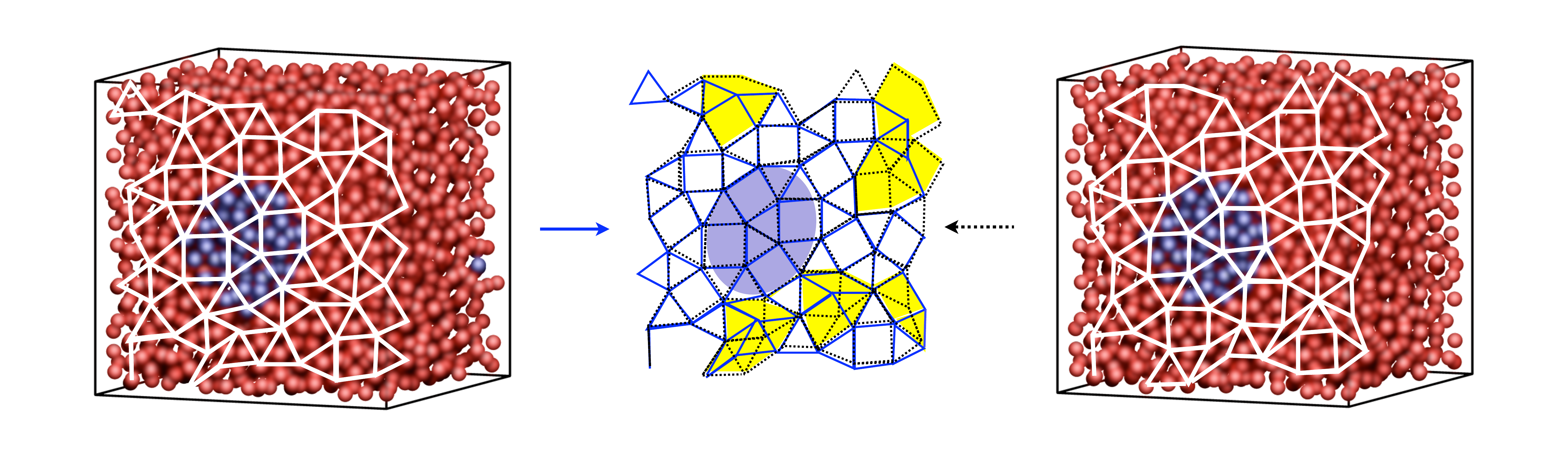}
    \end{minipage}\hfill
    \begin{minipage}[m]{0.29\linewidth}
\caption[Dependence of QC Tiling Arrangement on Liquid Structure]{Dependence of QC tiling arrangement on liquid structure.  The images show characteristic results from MC runs with the same quasicrystalline seed (blue) but with a different random number sequence.  At certain points in space, highlighted in yellow in the overlay, the tiling arrangements differ.}
\label{fig:aFigure2}
    \end{minipage}
  \end{center}
\end{figure*}

To obtain these results, we perform three distinct sets of computer simulations.  First, we use canonical (NVT) Monte Carlo (MC) to observe the growth of the QC from a static “seed” nucleus.  We then use isothermal-isobaric (NPT) MC to observe the growth of large QC nuclei, generated via umbrella sampling\cite{torrie77}.  Finally, we use umbrella sampling to generate many configurations containing nuclei to study the relationship between QC nuclei and icosahedral clusters.  All simulations contain 3375 atoms with pair interactions modeled via the Dzugutov potential\cite{dzug92}.  The form of the Dzugutov potential is identical to the 12-6 Lennard-Jones potential up to a distance at which an additional repulsive term dominates, suppressing the formation of BCC, FCC, and HCP crystals and favoring polytetrahedral ordering, where the 13-atom icosahedron is the ideal local structure.  In the Dzugutov supercooled liquid, atoms are known to organize into local energy-minimizing icosahedral clusters comprised of face-sharing and interpenetrating icosahedra\cite{doye03, dzug02, magnus}, which exhibit lower mobility than the bulk\cite{dzug02, magnus}.  The number of atoms that participate in icosahedral clusters at any time increases with the degree of supercooling\cite{zett01, magnus}.  At certain state points, the system forms a dodecagonal QC from the melt, which exhibits long-range polytetrahedral ordering\cite{dzug93, roth00} (see Fig.~\ref{fig:aFigure1}a).  Although the QC is physically stable over the timescale of a typical simulation, it is thermodynamically metastable with respect to the $\sigma$-phase periodic approximant\cite{roth00} (see Fig.~\ref{fig:aFigure1}b).  Here, we run simulations at temperature T=0.55, pressure P=3.5 and density $\rho$=0.85, which is slightly below the degree of supercooling ($T/T_m\sim0.7$) at which the system forms a QC in the absence of a seed nucleus or specialized simulation techniques.  At this state point, the growth of the solid phase occurs from a single nucleus, although under deeper supercooling many nuclei may grow simultaneously\cite{trudu06}.  

To observe the growing nucleus in our simulations, we define an order parameter to detect QC local ordering.  Our order parameter is a modification of the $\textbf{q}_6(i) \cdot \textbf{q}_6(j)$ scheme of reference~\cite{tenwolde96}.  There, the nearest-neighbor directions of an atom $i$ are expanded in spherical harmonics $Y_\ell(\theta, \phi)$ (with $\ell=6$) to construct a $2\ell+1$ dimensional complex vector $\textbf{q}_6(i)$, which can be thought of as a cluster ``shape-descriptor'' containing information regarding the shape and orientation of the cluster.  An atom \textit{i} forms a solid-like connection with neighbor \textit{j} if the vector dot product  $\textbf{q}_6(i) \cdot \textbf{q}_6(j)$ exceeds a certain value, and atoms with many solid-like connections are defined as being solid-like, reflecting the fact that in simple crystals all atoms have identical coordination shells.  This scheme must be modified for QCs and approximants, since neighboring atoms have non-identical coordination shells corresponding to different Frank-Kasper polyhedra\cite{frank58}.  For dodecagonal QCs, we increase the range of the neighbor cutoff to $r_{cut}=2.31\sigma$, corresponding to the first $\sim$2.5 neighbor shells.  Also, we modify the set of harmonics from $\ell=6$ to $\ell=12$, since we find that $\textbf{q}_{12}$ is sensitive to the symmetry of the dodecagonal QC, whereas $\textbf{q}_6$ produces no signal.  Pairs of atoms form a solid-like connection if $\textbf{q}_{12}(i) \cdot \textbf{q}_{12}(j) \geq 0.45$, with $\textbf{q}_{12}(i) \cdot \textbf{q}_{12}(j)$ normalized on the interval [0,1].  Atoms with $\geq 50\%$ solid-like connections are solid-like, otherwise they are liquid-like.  These cutoffs are chosen so as to maximize the distinction between liquid and QC; however, we note that the distinction becomes ambiguous near the liquid-solid interface where atoms exhibit properties that are intermediate between liquid and solid.  Therefore, for a diffuse nucleus, the solid-like atoms identified using this scheme represent only the nucleus core. 

We next define $q_6(t) \equiv \textbf{q}_6(i; t_0) \cdot \textbf{q}_6(i; t)$, autocorrelation function that measures how correlated atomic configurations are at time $\it t$ to their configurations at an earlier or later time $t_0$.  We base our scheme on $\textbf{q}_6$ rather than $\textbf{q}_{12}$, since our goal is to quantify how closely clusters match in terms of shape orientation, rather than to detect quasicrystalline correlations between non-identical neighbor shells.  We define $r_{cut}=1.65$ to include the first neighbor shell in our analysis.  We normalize $q_6(t)$ such that 1 is the maximum value and 0 represents the value for random correlations.  Configurations that are less correlated than the average random value have $q_6(t) < 0$. 

We begin by considering the growth of the solid phase from a small static seed nucleus in the form of a periodic approximant\cite{goldman93}  that is inserted into the MC simulation cell (see Fig.~\ref{fig:aFigure2}).   Approximants are crystals with identical local ordering to QCs; therefore, for small nuclei, QCs and approximants are identical and the difference in long-range ordering results from a different growth mechanism.  Constraining the seed in the form of an approximant allows us to determine whether the system requires a seed with a `special' structure to grow a QC.  We randomize our MC simulations at high temperature starting at time $t_r$ before quenching to T=0.55 at $t_q$, at which point atoms begin to attach to the seed, causing rapid solidification.  We observe that the system consistently forms a QC for all seed sizes, positions, and approximant structures, indicating that the system does not copy the seed, but rather incorporates atoms into the solid via a different paradigm.

Energy-driven QC growth models suggest that atomic attachment to the nucleus is deterministic, whereas entropy-driven models suggest that it is stochastic.  We test the applicability of these models for our system by modifying the random number sequence (RNS) used during the simulation, holding all else constant.  As depicted in Fig.~\ref{fig:aFigure2}, for the same seed nucleus (blue), we consistently obtain distinguishable QC tiling arrangements, indicating that QC growth has a stochastic element.  It is clear that the growth is energetically constrained as well, since most of the tiling discrepancies (yellow) represent ``phasons\cite{socolar86},'' tiling arrangements with nearly identical local energy.  Thus elements of both growth models appear relevant to QC growth.

Although the growth of the QC is affected by the RNS, the attachment of tiles to the nucleus is not random.  For random attachment, a change to the RNS causes an immediate change in the growth pathway, resulting in different tiling arrangements.  In contrast, our system exhibits an appreciable lag time between changes to the RNS and the appearance of tiling discrepancies.  For example, if we change the RNS at $t_q$, we observe fewer tiling discrepancies in the area immediately surrounding the nucleus than if we make a change at $t_r$.  (Note that in both cases, the nucleus is identical since the solid does not begin to grow until $t_q$).  This implies that QC growth is affected by stochasticity only insofar as it engenders differences in the local arrangement of atoms around the nucleus. 

We can test this idea quantitatively by using $q_6(t)$ to detect structural correlations between atoms surrounding the nucleus and the QC tiles that they subsequently form.  First, we generate many independent nucleation events in which the system grows a QC.  Previously, we used a seed to initiate nucleation; here we use umbrella sampling to generate many configurations with growing nuclei.  Our NPT MC runs are biased according to the harmonic weight function $w=\frac{1}{2}k\left( N-N_0 \right) ^2 $ \cite{tenwolde96, auer04}.  Here, $k=0.075$, $N$ is the number of atoms comprising the nucleus (measured by $\textbf{q}_{12}(i) \cdot \textbf{q}_{12}(j)$), and $N_0$ is specified such that nucleus sizes near $N_0$ are sampled selectively.  We slowly increase the bias from $N_0 = 10, 20, ... 90$ so that nuclei reach $N=80-100$.  We then use these microstates as starting points for unbiased NPT MC runs.  We observe that nuclei with $N>75$ atoms tend to grow, although factors other than size (e.g., shape, structure, etc.) may affect nucleus stability as well\cite{moroni05}.  We run MC simulations of growing nuclei for 75,000 MC cycles, the time it takes for nuclei to grow from $N\sim$100 to $N\sim$500.

\begin{figure}
\includegraphics[width=3.375in]{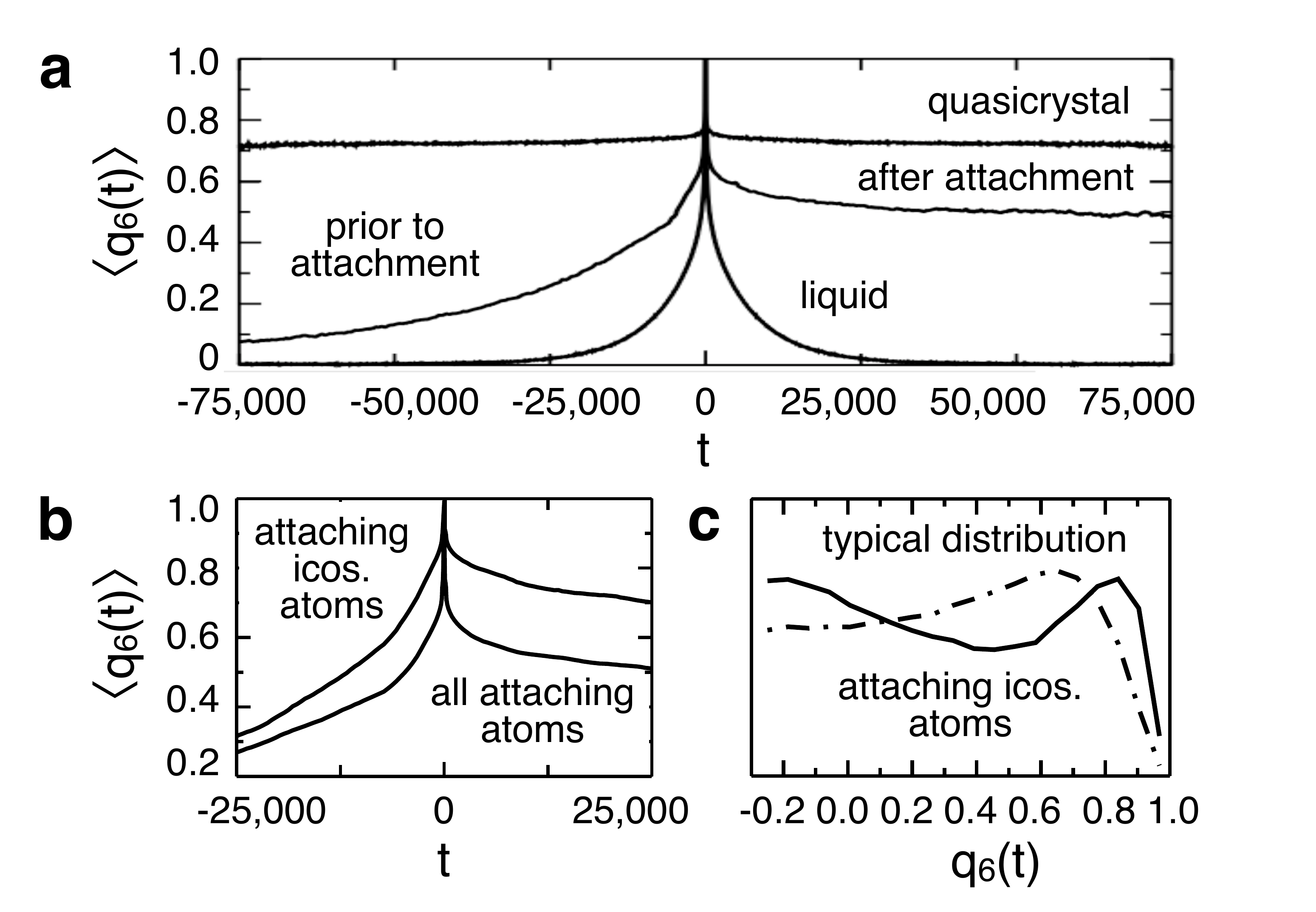}
\caption[Liquid-Solid Structural Correlations and Quasicrystal Growth]{Structural correlations. (a) Average value of $q_6(t)$ versus $t$ (MC steps).  From top to bottom: atoms in the dodecagonal QC, atoms in the non-equilibrium nucleating system that attach to the nucleus at $t=0$, atoms in the liquid.  For all runs, the reference time $t_0=0$. (b) Average value of $q_6(t)$ versus $t$ for attaching atoms.  Top: attaching atoms in icosahedral configurations.  Bottom: all attaching atoms.  (c)  Probability distribution of $q_6(t)$ at $<q_6(t) = 0.35>$.  Dotted line: the typical distribution for $<q_6(t) = 0.35>$, calculated from atoms in the supercooled liquid.  Solid line: attaching atoms in icosahedral configurations.}
\label{fig:aFigure3}
\end{figure}

We measure $\langle q_6(t) \rangle$ versus $t$ in the non-equilibrium nucleating system described above for atoms that attach to the growing QC nucleus at $t_0 = 0$, which we refer to hereafter as ``attaching atoms.''  For $t<0$, attaching atoms are in the region surrounding the nucleus, and for $t\geq0$, attaching atoms are in the solid nucleus (see Fig.~\ref{fig:aFigure3}a, middle curve).  We include only the atoms that attach permanently to the nucleus in our analysis, to ensure that we measure correlations between atoms in the QC and their former (non-solid) configurations rather than correlated reattachments of solid atoms.  Specifically, we exclude atoms that recross the 50\% threshold for solid-like connections (defined above) after fluctuations are averaged out.  Roughly 60\% to 70\% of the atoms attach without ever detaching.  

We compare $\langle q_6(t) \rangle$ for attaching atoms to atoms in the bulk QC and the bulk supercooled liquid at the same state point (Fig.~\ref{fig:aFigure3}a).  The value of $\langle q_6(t) \rangle$ is proportional to the degree of correlation to the reference structure at $t=0$.  This is exemplified by the high, constant value of $\langle q_6(t) \rangle$ observed for attaching atoms ($t>0$) and bulk QC atoms, which indicates a solid-like environment.  (The initial drop is due to thermal fluctuations).  For $t<0$, attaching atoms exhibit relatively high $\langle q_6(t) \rangle$, indicating that atoms joining the nucleus at $t=0$ are highly correlated to their former (pre-solidification) configurations.  

We can dissect the $\langle q_6(t) \rangle$ curve for attaching atoms into components based on local structure.  Overall, the dodecagonal QC consists of atoms in four different types of coordination shells:  icosahedral, Z13 , Z14, and Z15 configurations, where, `Zn' stands for the Frank-Kasper polyhedron\cite{frank58} with coordination number `n.'   We find that icosahedral atoms exhibit high $\langle q_6(t) \rangle$ (Fig.~\ref{fig:aFigure3}b), whereas other motifs do not deviate significantly from the average.  We rationalize the high value of $\langle q_6(t) \rangle$ for icosahedral atoms by considering the probability distribution of $q_6(t)$ at each point on the $\langle q_6(t) \rangle$ curve  (Fig.~\ref{fig:aFigure3}c).  We find that atoms in icosahedra, and, to a lesser extent, atoms in Z13 configurations (not shown), exhibit an unusually high proportion of strong correlations.  This indicates that as the nucleus grows, it incorporates a certain subset of icosahedral and Z13 atoms with minimal structural rearrangement.  Interestingly, Z14 atoms do not exhibit either high $\langle q_6(t) \rangle$ or a skewed $q_6(t)$ distribution, which indicates that although the icosahedral glass formed by the Dzugutov system has vibrational modes similar to the thermodynamically stable $\sigma$-phase\cite{simd00} (25\% icosahedra and 75\% Z14), the most correlated atoms do not exhibit $\sigma$-like character.  Rather, the high degree of icosahedrality and the presence of correlated Z13 atoms (which do not appear in the approximants but are highly present in the supercooled liquid) indicate that atoms in liquid-like icosahedral clusters surrounding the nucleus tend to retain their configurations during incorporation into the nucleus.  

We can obtain a more intuitive picture of the role of icosahedral clusters by considering their spatial arrangement in relation to the growing QC nucleus.  We generate a large number of nuclei using the umbrella sampling scheme outlined above.  To expedite sampling, we allow configuration swapping between simulations via parallel tempering\cite{auer04}.  In all, we run 10 simultaneous MC simulations for 3.5 million MC steps, where each simulation has a unique biasing potential minimum $N_0=10,20, \ldots100$ for a given simulation.  We save configurations every 100 MC steps, giving us 35,000 total microstates containing nuclei of sizes $N$=10-110 for analysis.  We identify icosahedral clusters in our microstates using the method of reference\cite{iac07}, an extension of the method of reference\cite{stein83}.  

\begin{figure}
\includegraphics[width=3.375in]{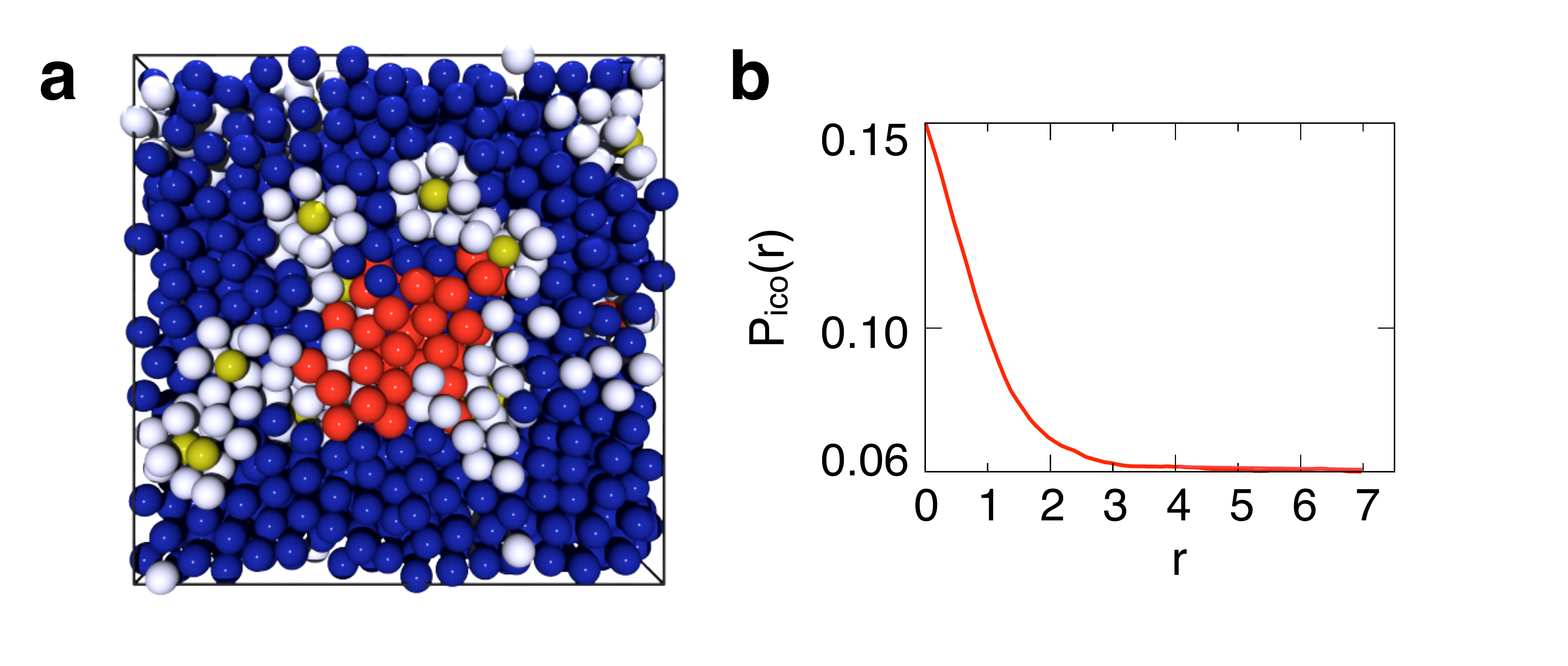}
\caption[Icosahedral Wetting of Quasicrystal Nucleus]{Icosahedral environment. (a) Simulation snapshot showing a QC nucleus (red) together with icosahedral clusters (yellow for icosahedral centers, white for surface atoms) in the liquid (blue). (b) The average probability of observing an atom at the center of  an icosahedron versus r, the distance from the nucleus surface.}
\label{fig:aFigure4}
\end{figure}

As depicted in Fig.~\ref{fig:aFigure4}a, we find that icosahedral clusters (yellow, white) ``wet'' the core of the QC nucleus (red), a mechanism that may reduce interfacial tension\cite{kelton91}.  We quantify the tendency for icosahedral clusters to aggregate around the nucleus by calculating $P_{ico}(r)$, the average probability of observing an atom at the center of an icosahedron a distance \textit{r} away from the nucleus surface (see Fig.~\ref{fig:aFigure4}b).  For nuclei of all sizes, we observe that $P_{ico}(r)$ starts with a value of 0.15 near the nucleus surface and decreases to the liquid value of 0.06 over a range of about three particle diameters, indicating that there is an increased presence of icosahedral clusters in the region surrounding the nucleus.  As the nucleus grows, it must change the connectivity of these clusters from liquid-like local-energy minimizing arrangements to ordered quasicrystalline arrangements.  The tendency to retain the configurations of some of the clusters rather than copying the nucleus surface template is the ``growth rule'' underlying the formation of the QC.
 
Our results demonstrate how QCs provide a `path of least resistance' for solid phase growth versus crystals.  In this case, whereas the stable $\sigma$-phase approximant must rearrange kinetically trapped atoms into a crystal lattice, the less constrained QC is able to reach a `structural compromise' with the surrounding atoms to grow more rapidly.  Our results explain why QCs often form in rapidly quenched metallic alloys, as these systems produce rapidly growing nuclei as well as low-energy icosahedral clusters.  In terms of QC growth models, our results give physical insight into how the nucleus `decides' to form a particular tile as it grows.  We note that although icosahedral clusters are not the energy-minimizing structural motif for all QCs, the basic mechanism at hand -- the tendency for certain atoms to retain their liquid configuration when incorporated into the growing solid nucleus -- should hold generally for QC-forming systems.

\textbf{\textit{Acknowledgements:}} We thank D. Frenkel and A. Cacciuto for assistance with umbrella sampling.  We also thank M.N. Bergroth and J. Mukherjee.  Funding provided by NASA (DE-FG02-02ER46000) and DoEd (GAANN).
\bibliography{qc}

\end{document}